\documentclass[10pt]{article}
\usepackage{amsmath}
\usepackage{graphicx}
\usepackage[labelfont=bf]{caption}
\usepackage{hyperref}
\usepackage{cite}
\usepackage{geometry}
\usepackage{lipsum}
\usepackage{setspace}
\usepackage{caption}
\usepackage{subcaption}
\usepackage{amssymb}
\usepackage{longtable}
\usepackage[dvipsnames]{xcolor}
\usepackage{multicol}

\definecolor{myred}{RGB}{0, 0, 0}

\begin{document}

\bibliographystyle{naturemag}

\title{Geometric percolation of colloids in shear flow}
\author{Ilian Pihlajamaa\footnote{Corresponding author. Email: url{i.l.pihlajamaa@tue.nl}}\ \footnote{Group of Soft Matter and Biological Physics, Eindhoven University of Technology, De Groene Loper 19, 5612 AP Eindhoven, The Netherlands.} , René de Bruijn$^\dag$, Paul van der Schoot$^\dag$}
\maketitle

\begin{abstract}
We combine a heuristic theory of geometric percolation and the Smoluchowski theory of colloid dynamics to predict the impact of shear flow on the percolation threshold of hard spherical colloidal particles, and verify our findings by means of molecular dynamics simulations. It appears that the impact of shear flow is subtle and highly non-trivial, even in the absence of hydrodynamic interactions between the particles. The presence of shear flow can both increase and decrease the percolation threshold, depending on the criterion used for determining whether or not two particles are connected and on the P\'{e}clet number. Our approach opens up a route to quantitatively predict the percolation threshold in nanocomposite materials that, as a rule, are produced under non-equilibrium conditions, making comparison with equilibrium percolation theory tenuous. Our theory can be adapted straightforwardly for application in other types of flow field, and particles of different shape or interacting via other than hard-core potentials.
\end{abstract}
\begin{multicols}{2}

\section*{Introduction}\label{sec1}

The electrical conductivity of polymeric materials can be varied over ten orders of magnitude by the incorporation of a relatively small fraction of conductive nano-fillers, such as carbon black, graphene and metallic particles \cite{li2005nanohybrid, moniruzzaman2006polymer, wu2008preparation, kim2009strategy, king2010electrical, hornbostel2006single, satapathy2007tough, abbasi2009rheological, satapathy2007tough,stankovich2006graphene}. The intense interest in this topic, evidenced by a huge surge in the number of studies dealing with polymeric nano-composite materials in the last few decades, is perhaps not entirely surprising given their potential technological applications in, say, opto-electronics, photo-voltaics and electromagnetic interference shielding \cite{hermantmanipulating, zeng2016thin, avouris2008carbon, kymakis2002single, kymakis2003high}.
It turns out that the degree of homogeneity of the nano-particle dispersion in the host material is of crucial importance to the level of conduction of the composite achieved \cite{sumita1991dispersion, choudhary2011polymer, li2007high, ramasubramaniam2003homogeneous, mathieu2006processing, huang2014electrical, ou2003assessment, majidian2017role}. Hence, great care is taken in the manufacturing process to disperse the nano-fillers evenly when the host material is still in the fluid stages of the production. For this purpose, techniques are applied that include sonication, manual mixing and shear mixing, followed by casting and curing of the composite \cite{kang2013comparison, spitalsky2010carbon, tiarks2001encapsulation, cotten1984mixing, rwei1992analysis}. These methods contribute to the homogeneous dispersion of the particles and prevent their aggregation. Because the curing of the fluid is typically (but not always \cite{grossiord2008influence}) done relatively quickly after the mixing so as to avoid re-aggregation, the out-of-equilibrium structure of the nano-fillers should be expected to be essentially frozen-in in the final, solid composite.

Consequently, if one attempts to predict the percolation threshold of composite materials using the standard tools of liquid state theory, as is usually done in connectedness percolation theory \cite{torquato2002random}, then it stands to reason that these predictions must be flawed. Indeed, connectedness percolation theory assumes the particle distribution to obey (equilibrium) Boltzmann statistics \cite{coniglio1977pair, xu1988analytic}. To remedy this for those conditions where the particle distribution does not obey equilibrium statistics, a quantitative, out-of-equilibrium continuum percolation theory is sorely needed. Unfortunately, no such theory is, as far as we are aware, currently available. One could envisage setting up a non-equilibrium version of connectedness percolation theory, based on the Smoluchowski equation for the steady-state pair correlation function under flow. \cite{dhont1989distortion} However, in the dynamical theory there is no obvious way of separating the non-equilibrium equivalent of the so-called connectedness and blocking functions, as is possible in thermodynamic equilibrium \cite{coniglio1977pair}.

An alternative that does not have this problem, is to take a more heuristic approach such as that we put forward in this work. 
Here, we use a simple \textit{geometric} criterion for the percolation threshold that depends solely on the pair correlation function and a connectivity criterion. The method quantitatively describes results for different particle shapes from computer simulations under conditions of thermal equilibrium \cite{alon1991new}. We combine this with the known steady-state solution of the Smoluchowski equation for the pair correlation function in shear flow, that we solve in the limit of low volume fractions but apply also to intermediate and high concentrations \cite{blawzdziewicz1993structure}. Together, these two ingredients provide us with a simple and tractable way to predict the percolation threshold in dispersions that are out-of-equilibrium, which we apply to the case of colloidal hard spheres in simple shear flow. We note that the method is quite generic and allows for a straightforward extension to other types of flow and other types of inter-particle interaction and particle shape. Also, since we focus attention on steady-state flow, our theory applies not only to particles in simple fluids but also in (visco-elastic) polymeric ones.

Simple shear flow, parametrised by the velocity field $\textbf{v}~\propto~(y, 0, 0)$, is among the most studied flow fields due to it being the simplest model for the more complex shear flows that are ubiquitous in industrial and experimental processes. Moreover, it is well known that the properties of colloidal materials change drastically when subjected to such flows \cite{bergenholtz2001theory, vermant2005flow}. Specifically, both shear-induced cluster formation and breaking-up has been reported in the literature \cite{butler1999shear, koumakis2015tuning, gallier2015percolation,de1979conjectures}, making its influence on the percolation threshold highly unclear. With this work we aim to shed light on what impact a shear flow field may have on connected clusters, and in particular under what conditions these break up or grow to give rise to percolating particle networks.

\begin{figure*}[ht]
\centering
\includegraphics[width=\linewidth]{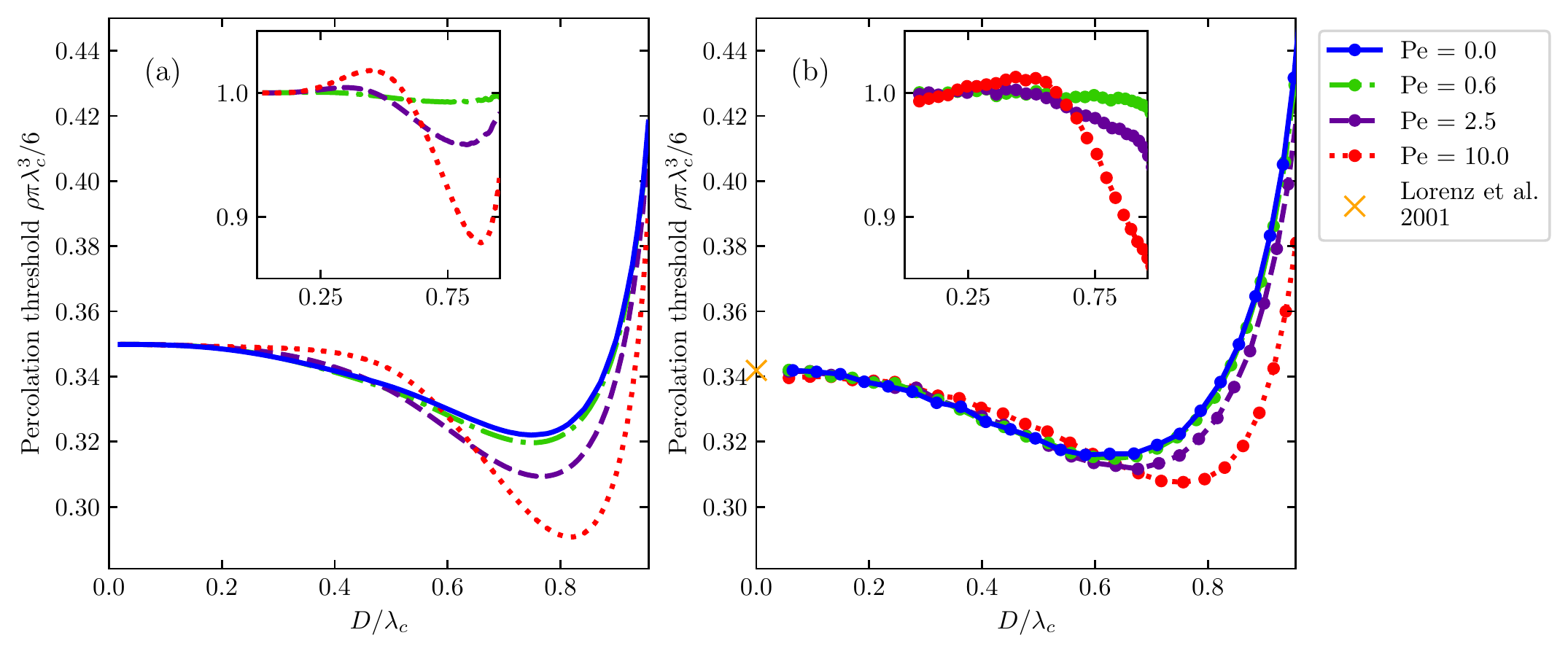}
\caption{Theoretical (a) and simulation (b) results of the dimensionless percolation threshold of a dispersion of hard spherical particles subject to a simple shear flow as function of the ratio of the hard core diameter $D$ and connectivity length $\lambda$. The strength of the shear flow is quantified with the P\'{e}clet number $\mathrm{Pe}$. In (b), we indicate with the cross the literature value of $\rho\pi\lambda_c^3/6=0.341889$ valid in the case of non-interacting particles \cite{lorenz2001precise} and we add lines between the obtained data points as guides to the eye. The insets show the relative change of the percolation threshold compared to the no-flow case. In all figures, the maximal value of the hard-core diameter  is given by $D/\lambda_c=0.96$.}
\label{fig:percflow}
\end{figure*}

As we shall see, while our theory loses its quantitative nature for high shear rates when compared with our Langevin dynamics simulations, it does overall describe the full phenomenology of how the percolation threshold depends on the connectivity range, that is, the geometric criterion defining whether two neighbouring particles are connected or not. This agreement is reached without including any free (fitting) parameters. In agreement with our simulations, we find that for sufficiently large connectivity range, fluid flow increases the percolation threshold and more so the larger the P\'{e}clet number, whilst for small connectivity range the opposite happens. This is caused by the delicate balance between the compression and extension that the flow field exerts on the pair structure. The impact of fluid flow on the percolation threshold remains modest, however, staying within 15\% of the static case for P\'{e}clet numbers up to about ten. Our predictions are summarised in Figs.~\ref{fig:percflow}a and the results of our simulations in Fig.~\ref{fig:percflow}b.

In the following, we first summarise the ingredients of our theory, consisting of the geometric percolation theory of Alon and collaborators \cite{alon1991new}, and the analytical prediction of the correction of the pair structure by shear flow of B\l{}awzdziewicz \textit{et al.} \cite{blawzdziewicz1993structure}. The latter we compare with our Langevin dynamics simulations. We subsequently integrate both ingredients and obtain a prediction of the percolation threshold, and end the paper with conclusions and an outlook. Details of our calculations and simulations are given at the end of this paper. 

\section*{A heuristic approach to percolation}\label{sec2}

The theoretical framework of Alon \textit{et al.} for predicting the percolation threshold is ideally suited for our problem of percolation far out of equilibrium, as it is geometric in nature and does not require thermodynamic equilibrium to hold \cite{alon1991new}. We summarise it below for the case of spherical interacting particles but note that our treatment can be extended to non-spherical particles as well, see Ref. \cite{thovert2017percolation} for an example in the case of non-interacting particles.

The theory presumes $N$ homogeneously dispersed particles to be present in a volume $V$. We quantify the structure of this dispersion with the pair correlation function $g(\textbf{r})$: if a particle is placed at the origin, then the probability of finding another particle in volume element $\mathrm{d}^{3}\textbf{r}$ at position $\textbf{r}$ is equal to $\rho g(\textbf{r})\mathrm{d}^{3}\textbf{r}$, where $\rho=N/V$ is the number density. We further assume that pairs of particles with a centre-to-centre distance $|\textbf{r}|$ smaller than the connectivity length $\lambda$ are directly connected to each other, implying that charge carriers can be transported efficiently between them. Physically, this connectivity range $\lambda$ can \textit{e.g.} be interpreted as a tunnelling length \cite{hu2006hopping,kyrylyuk2008continuum}. 

The question arises how to find the smallest connectivity length $\lambda_c$ for which a connected cluster of particles exists that spans the entire material for a given density $\rho$. This is equivalent to asking what the critical density $\rho_c$ is at which a percolating cluster appears for a given connectivity length $\lambda$ \cite{ambrosetti2010solution, nigro2011transport}.  The advantage of the former formulation is that finding the structure for a given density and finding the percolation threshold for a given structure are now completely decoupled problems, and no longer have to be solved self-consistently. 

The argument of the theory is as follows. Clearly, a macroscopically connected cluster must comprise many so-called backbone particles with two or more connections, since particles with one or no connections at all cannot propagate connectivity and therefore do not contribute to percolation. Heuristically, Alon and coworkers argue that percolating networks only exist if the average distance $L$ between such backbone particles is smaller than twice the average distance $l$ between directly connected particles (for which $|\textbf{r}| < \lambda$). The percolation threshold may thus be found by requiring that $L=2l$ \cite{alon1991new}. 
Estimates for both lengths $l$ and $L$ can be obtained relatively straightforwardly from the pair correlation function, $g(\textbf{r})$. 

To start with the first and recalling that the probability of finding a particle in volume $\mathrm{d}^3\textbf{r}$ at position $\textbf{r}$ is $\rho g(\textbf{r})\mathrm{d}^3\textbf{r}$, we can use an appropriately normalised (statistical) moment of $g(\textbf{r})$, evaluated within the connectivity region, to estimate the mean distance $l$ between connected particles: $l^2=\int_{V_\lambda} \mathrm{d}^3\textbf{r} g(\textbf{r})r^2 / \int_{V_\lambda} \mathrm{d}^3\textbf{r} g(\textbf{r})$. We use the second moment rather than the first because it yields more accurate predictions for the percolation threshold \cite{alon1991new}. In the expression for $l^2$, $V_\lambda$ is the connectivity region, which in our case is the region within the sphere of radius $\lambda$.

As to the distance $L$ between backbone particles with two or more direct connections, it is convenient to assume that the probability $P_k$ that a particle has $k$ direct connections is Poissonian, \textit{i.e.}, $P_k = B^k \exp(-B)/k!$, in which $B=\rho \int_{V_\lambda} \mathrm{d}^3\textbf{r}g(\textbf{r})$ is the average number of direct connections. This assumption is exact for non-interacting particles \cite{wax1954selected, reichl1999modern}, and from our molecular dynamics simulations we find that it remains a good approximation for dispersions of hard spheres provided the density is not near close packing (results shown in Fig.~\ref{fig:fullplotsimlines}(g)). 
The mean number density of particles with at least two neighbours, $\rho_2$, can now straightforwardly be found from $\rho_2 = \rho \left(1-P_0-P_1\right) = \rho \left(1-(1+B)\exp(-B)\right)$. Assuming that the volume available to each particle is spherical, we have $L=2(4\pi\rho_2/3)^{-1/3}$. 

In conclusion, the criterion for percolation, $L=2l$, we find to be a function only of $\rho$, $g(\textbf{r})$, and $\lambda$, and for any given density $\rho$ and corresponding pair correlation function $g(\textbf{r})$, we can use known numerical routines \cite{brent2013algorithms} to solve for the percolation threshold $\lambda_c$. Of course, we need to know $g(\textbf{r})$ too. It can be evaluated directly either from computer simulations, or, say, from the (numerical) integration of the Ornstein-Zernike equations with a suitable closure. For the case of hard particles, the Percus-Yevick closure is known to be highly accurate \cite{hansen}.

In Fig.~\ref{fig:percflow}a and Fig.~\ref{fig:percflow}b we also show our theoretical and simulation results for zero P\'{e}clet number, so in the absence of a shear field, the former within Percus-Yevick theory. For this, we numerically integrated the Ornstein-Zernike equation within the Percus-Yevick closure \cite{homeier1995iterative}.
The difference between theory and simulation is less than 6\% for all values of the connectivity length $\lambda$, which incidentally outperforms approximations derived from rigorous liquid state theory significantly \cite{desimone1986theory}. Notice that the limit $D/\lambda \rightarrow 0$ corresponds to the penetrable-sphere limit. For increasing values of $D$ at fixed value of $\lambda$, $D/\lambda$ increases and the percolation threshold initially decreases before it increases again. This is caused by the competition of an increasing contact value of the pair correlation function and a shrinking connectivity region. As we shall see, the presence of a flow field alters that competition. To study this, we first need to evaluate the impact of flow on the pair structure. A brief discussion of this we provide in the next section.

\section*{Pair correlations under shear flow}\label{sec3}

In order to find the percolation threshold in particle suspensions subject to a flow field, we must take into account the impact of the flow field on the pair correlation function, $g(\textbf{r})$. To do this, we approximate the pair correlation function using a steady-state two-particle Smoluchowski equation.

In the absence of hydrodynamic interactions and many-body correlations, this two-body Smoluchowski equation reads \cite{dhont1989distortion}
\begin{equation}\label{eq:smol}
    \nabla \cdot \left( \Gamma \textbf{r} g - 2D_0\left(g\nabla V/k_BT+\nabla g \right) \right) = 0,
\end{equation}
for an arbitrary velocity gradient tensor $\Gamma$ and spherically symmetric interaction potential $V/k_BT$. Here, we introduced the thermal energy $k_BT$ to non-dimensionalise the pair potential.

\begin{figure*}[ht]
    \centering
    \includegraphics[width=\textwidth]{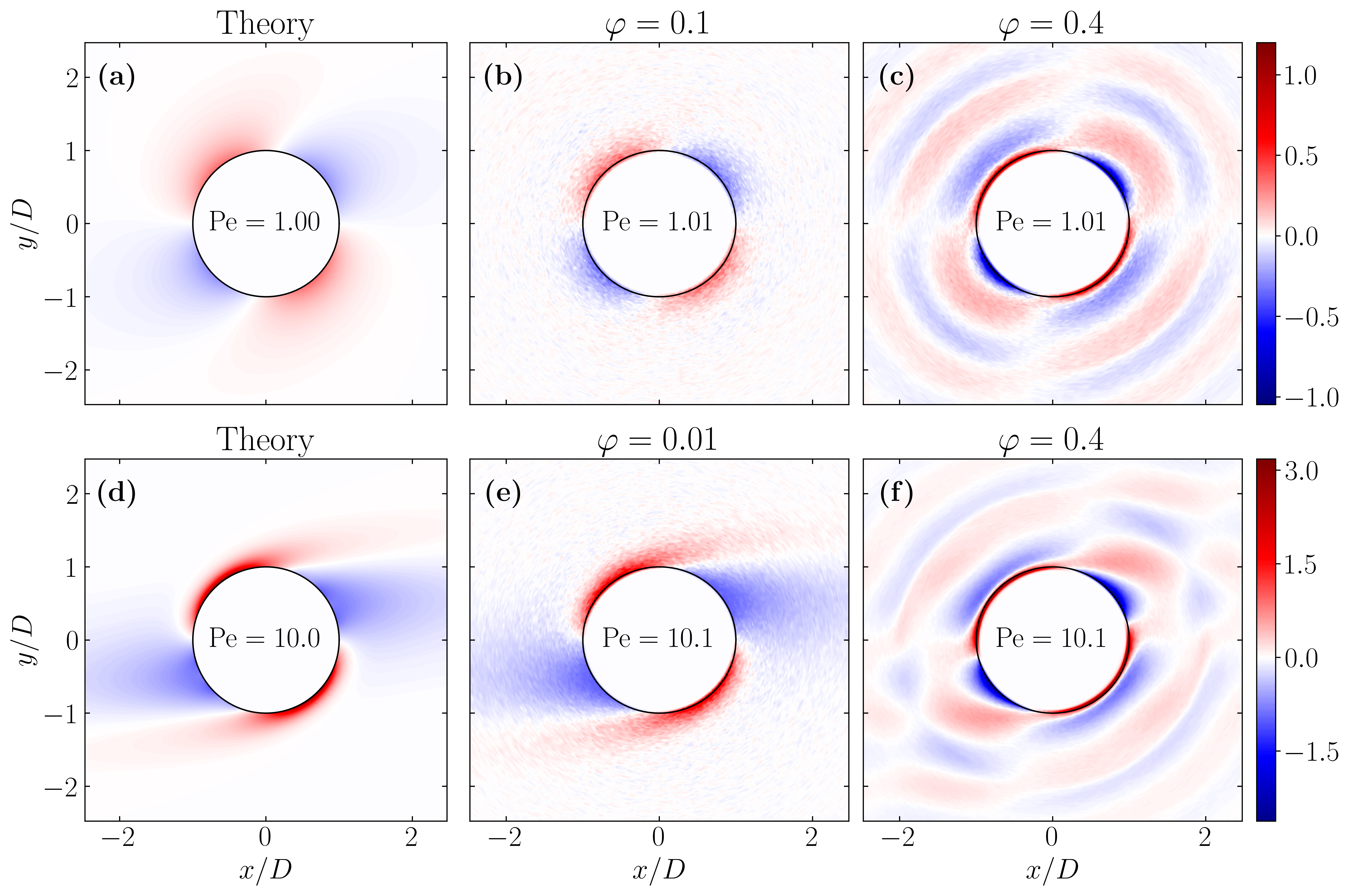}
    \caption{Comparison between theory and simulations of the shear-induced correction of the pair correlation function $\delta g(\textbf{r}, \mathrm{Pe})$ at low and high volume fractions of hard, spherical particles in the flow-gradient plane. The figures in the first column (a, d) correspond to the theoretical model, whereas the second (b, e) and third (c, f) column correspond to simulation results at low and high volume fractions $\varphi$. The first row (a-c) represents  dispersions at P\'{e}clet number $\mathrm{Pe}=1$, and for the second we set $\mathrm{Pe}=10$. Each row shares a colour scheme that indicates the values of $\delta g(\textbf{r}, \mathrm{Pe})$. 
    For clarity, we added a thin black line at $r=D$ indicating the theoretical excluded volume of hard, spherical particles}
    \label{fig:fullplotsim}
\end{figure*}

We restrict ourselves to the case of simple shear flow, with velocity field $\textbf{v} = \Gamma \textbf{r} = \dot\gamma y \hat{\textbf{x}}$, where $\dot\gamma$ is the shear rate, $\hat{\textbf{x}}$ the unit vector in the flow direction and $y$ the coordinate along gradient direction of the flow field. To calculate the pair structure in such simply sheared liquids, we view it as a (not necessarily small) correction $\delta g(\textbf{r}, \mathrm{Pe}) = g(\textbf{r}, \mathrm{Pe}) - g_0(\textbf{r})$ of the equilibrium pair correlation function $g_0(\textbf{r})$. Here, $\mathrm{Pe} = \dot\gamma D^2/4D_0$ is the P\'{e}clet number, with $D$ the particle diameter and $D_0$ the self-diffusion constant\footnote{By self-diffusion constant, we mean the diffusion constant that a particle would have if it is isolated from all other solute particles in the absence of a flow field. It can be estimated by the Stokes-Einstein relation $D_0=k_BT/3\pi\eta_0D$, in which $\eta_0$ is the viscosity of the medium.}. For spherical particles, the equilibrium pair correlation function $g_0(\textbf{r})$ only depends on the radial distance $r=|\textbf{r}|$ between the particles, for which reason it is usually referred to as the radial distribution function. Under shear, this is no longer true as the flow field breaks the spherical symmetry. We neglect the effect of hydrodynamic interactions to keep the theory concise. At the end of the next section, we shall discuss the validity of this approximation.

To apply \eqref{eq:smol} to the case of a hard-sphere liquid,  B\l{}awzdziewicz and Szamel \cite{blawzdziewicz1993structure} have imposed no-flux boundary conditions at $r=D$ and set $V=0$ for $r>D$. This results in the boundary value problem for $\delta g(\textbf{r}, \mathrm{Pe})$ given by
\begin{align}
    \label{pde}2D_0\nabla^2 \delta g - \dot\gamma y \frac{\partial\delta g}{\partial x} &= 0 &r>D,\\
    \label{bcs}
    2D_0 \frac{\partial \delta g}{\partial r} - \dot\gamma \frac{xy}{r} \delta g &=\dot\gamma \frac{xy}{r} &r=D.
\end{align}
Since many-body correlations have been neglected here, the prediction for $\delta g(\textbf{r})$ does not depend on the particle density. This approximation is equivalent to setting the reference pair correlation function $g_0(r)=\exp(-V(r)/k_BT)$. We therefore expect the predicted flow-induced correction to be accurate only for sufficiently low volume fractions. In the next section, we combine this correction, strictly valid in the dilute limit, with a more realistic equilibrium structure at high densities.

We choose to use the theory of B\l{}awzdziewicz and Szamel  \cite{blawzdziewicz1993structure}, rather than more sophisticated approaches \cite{abe1959kirkwood, dhont1994effect, ohtsuki1981dynamical,rice1965equation, nazockdast2012microstructural,schwarzl1986shear, szamel2001nonequilibrium,banetta2019radial,nazockdast2012microstructural, wagner1989nonequilibrium},
because even with this relatively simple model we obtain quite accurate results for the percolation threshold. An additional benefit is that the partial differential equation \eqref{pde}, together with hard-sphere no-flux boundary condition \eqref{bcs}, admit an analytical treatment that we summarise in Materials and Methods \cite{blawzdziewicz1993structure}.

We perform a direct test of the theory by comparing the prediction for $\delta g(\textbf{r}, \mathrm{Pe})$ with results from our Langevin dynamics simulations in Fig. \ref{fig:fullplotsim}. The figure shows the shear-induced correction $\delta g$ obtained from the theory and that from our simulations at two densities in the $xy$-plane, for two different P\'{e}clet numbers. We see that for low volume fractions $\varphi = \pi \rho D^3/6 \ll 1$, the simulation results quantitatively match the theoretical model, at least up to $\mathrm{Pe}=10$.
(For $\mathrm{Pe}=1$ we take for our simulations $\varphi=0.1$ rather than the $\varphi=0.01$ that we used for $\mathrm{Pe}=10$ to improve the statistics.)
For high volume fractions, however, we find that highly complex and long-ranged pair correlations are induced by the shear field, which differ not only quantitatively but in fact also qualitatively from those that our theoretical model predicts. This is not all that surprising, given the approximations of the model.

\begin{figure*}[t]
    \centering
    \includegraphics[width=0.95\linewidth]{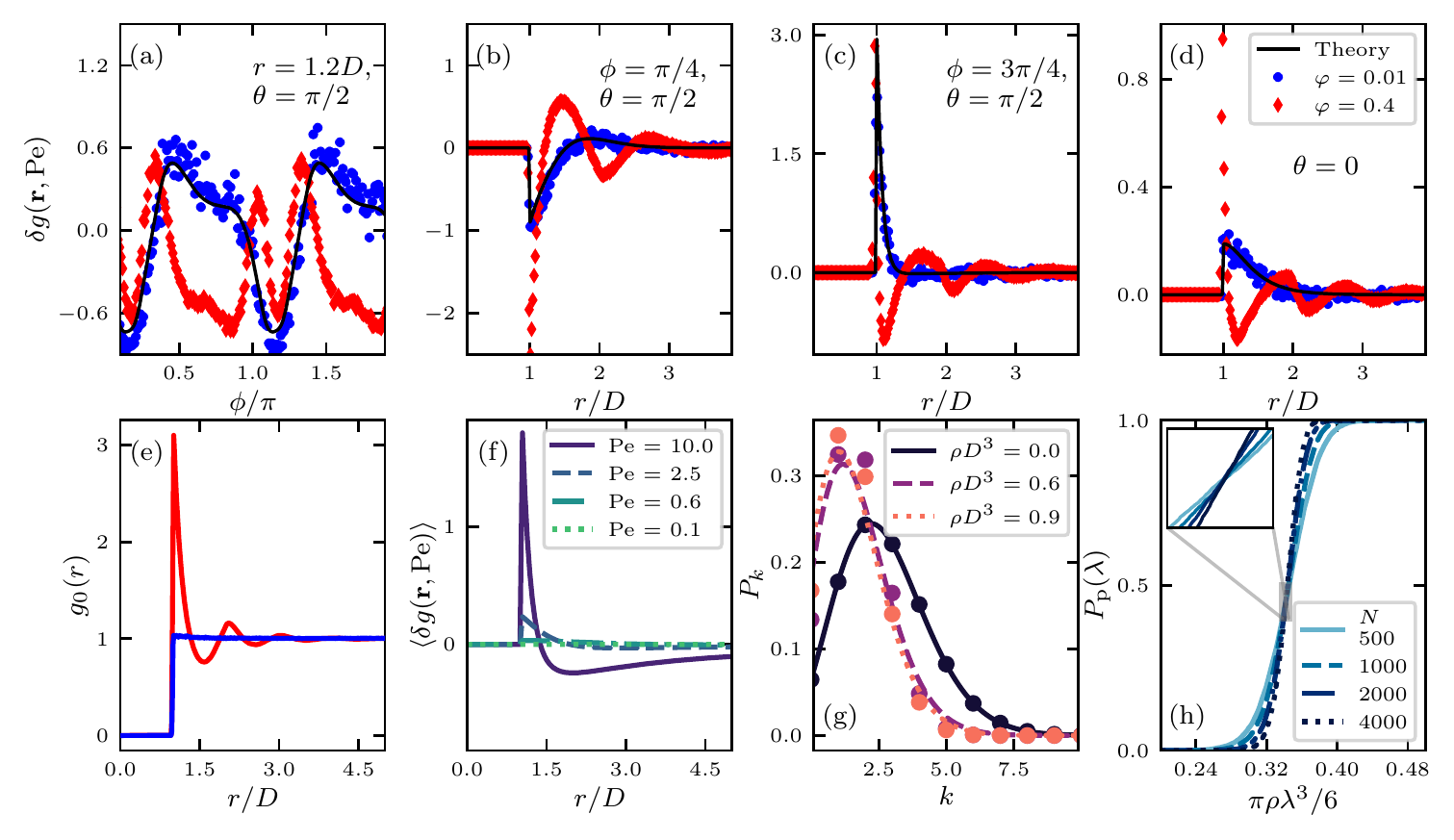}
    \caption{Quantitative comparison of the predicted structure with molecular dynamics simulation data. (a-d) Shear-induced correction of the pair correlation function $\delta g(\textbf{r},\mathrm{Pe})$ of spherical particles along different curves through three dimensional space at volume fraction $\varphi=0.01$ (blue circles) and $\varphi=0.4$ (red diamonds) for P\'{e}clet number $\mathrm{Pe}=10$. Respectively, they correspond to the correction at constant radial distance $r=1.2D$ in the $xy$-plane, and along the curves $\textbf{r} = r(\hat{\textbf{x}}+\hat{\textbf{y}} )/\sqrt{2}$, $\textbf{r} = r(-\hat{\textbf{x}}+\hat{\textbf{y}} )/\sqrt{2}$, and $\textbf{r} = r\hat{\textbf{z}}$. Indicated are also the theoretical predictions (drawn line). (e) Equilibrium radial distribution function at volume fractions $\varphi=0.01$ (blue) and $\varphi=0.4$ (red). (f) Flow-induced correction of the pair correlation function averaged over all solid angles for different P\'eclet numbers as predicted by the theory. Colour coding of the curves is given in the legend. (g) Probability $P_k$ that a particle has $k$ direct connections at the percolation threshold $\lambda_c$ in the absence of shear flow. The simulation results are compared to Poisson distributions where the average number of neighbours $B$ is determined from the simulation results, that is, $B=2.7, 1.7, 1.6$ for $\rho D^3=0.0, 0.6, 0.9$, respectively. (h) Percolation probability $P_\mathrm{p}(\lambda)$ in the absence of shear flow for $D/\lambda=0$, determined for different particle number $N$ indicated in the legend. The
    intersection of the curves is
    shown in the inset.
    }
    \label{fig:fullplotsimlines}
\end{figure*}

At high densities, long-ranged correlations are in fact also present in the equilibrium pair correlation function, shown in Fig.~\ref{fig:fullplotsimlines}(e), to which the shear flow couples. At low P\'{e}clet numbers, the peaks of the equilibrium pair correlation function are amplified in the compression quadrants and suppressed extensional quadrants. The opposite happens with the troughs of the equilibrium radial distribution function. In the compression quadrants, the flow increases the magnitude of structural correlations because the local density increases, whereas the opposite happens in the extensional quadrants. If the strength of the shear flow increases, this persists in the compression quadrant but leads to complex structural patterns in the extensional quadrant.

A more quantitative comparison we provide in Fig. \ref{fig:fullplotsimlines}(a--d), where we plot the same results for $\delta g(\textbf{r},\mathrm{Pe})$ presented in Fig. \ref{fig:fullplotsim}(d--f) for $\mathrm{Pe}=10$ along specific curves. In Fig. \ref{fig:fullplotsimlines}(a) we vary the azimuthal angle $\phi = \arctan{y/x}$ at fixed distance in the $xy$-plane and in (b--d) the radial distance $r$ for fixed angles $\theta$ and $\phi$. 

The figures confirm that the theoretical model reproduces very well the shear-induced correction of the pair correlation function $\delta g$ at low volume fraction. At high volume fractions of particles, we again see that the theoretical model qualitatively fails to predict the features of $\delta g(\textbf{r}, \mathrm{Pe})$. In fact, we notice the emergence of an additional peak in the correction of the pair correlation function at constant radius, see Fig.~\ref{fig:fullplotsimlines}(a). In the constant-angle plots, (b--d), we obtain additional peaks and a large increase in the structural correlations at high particle density. In Fig.~\ref{fig:fullplotsimlines}(d), we show the $\delta g$ along the line perpendicular to the shear-plane. This correction is clearly nonzero, which is consistent with earlier findings from theory and simulations \cite{morris2002microstructure, nazockdast2012microstructural, nazockdast2013pair, blawzdziewicz1993structure}. Notice also that the \textit{range} of these correlations increases drastically at a volume fraction of $\phi = 0.4$, which is close to the freezing transition in the absence of a flow field. 

What these results point at, is that in the $xy$-plane along the axis $\phi=\pi/4$ (in the extensional quadrants) the contact value of $g(\textbf{r})$ goes down, whilst along the axis $\phi=3\pi/4$ (in the compressional quadrants) it increases very strongly. This, of course, has consequences for the structure, size and shape of (geometric) clusters of particles, as the contact value of $g(\textbf{r})$ informs on the likelihood of the presence of other particles near the surface of a test particle. In turn, this must have an impact on the percolation transition, which occurs when macroscopic clusters form. We investigate this next in more detail by making use of the prediction for $\delta g(\textbf{r}, \mathrm{Pe})$ and the heuristic percolation criterion to investigate the percolation threshold for different P\'{e}clet numbers.

\section*{The percolation threshold}\label{sec4}

In the previous section we calculated the correction $\delta g(\textbf{r}, \mathrm{Pe}) = g(\textbf{r},\mathrm{Pe}) - g_0(r)$ of the equilibrium radial distribution function $g_0(r)$, if we presume the latter to obey a simple Boltzmann weight. For hard particles this becomes a step function, and is accurate only for very low volume fractions $\varphi \ll1$. Since we are not necessarily only interested in results at low volume fractions, we use the $g_0(r)$ obtained from the numerical integration of the Ornstein-Zernike equation together with the Percus-Yevick closure \cite{homeier1995iterative} to obtain an \textit{ad hoc} approximation for the full multi-body $g(\textbf{r},\mathrm{Pe})$. Clearly, it is only multi-body for the (reference) equilibrium structure, while the impact of flow is only described at the two-body level.
Inserting this into the percolation criterion yields the percolation threshold, which we present in Fig. \ref{fig:percflow}.

We show the results on the effect of shear flow on the percolation threshold according to our model in Fig. \ref{fig:percflow}a. In Fig. \ref{fig:percflow}b, we compare them to percolation thresholds obtained from the analysis of snapshots of our molecular dynamics simulations, on which more information can be found in the Materials and Methods section.  We plot the critical volume fraction $\rho\pi\lambda_c/6$ of connectivity regions as a function of the relative size of the hard-core particle diameters $D/\lambda_c$. If the hard-core diameter $D$ goes to zero and particle interactions become negligible, we find that the effect of the shear flow on the percolation threshold does so too. In this case, our simulation results agree very accurately with the literature value of $\rho\pi\lambda_c/6 \approx 0.341 889$.\cite{lorenz2001precise} The fact that a flow field does not impact this number is intuitive, since the flow field cannot induce any structural changes if the particles do not interact, \textit{i.e.}, if $D=0$. In the intermediate and high $D/\lambda_c$ regime, both theory and simulations agree that an applied shear flow can both increase and decrease the percolation threshold of a fluid dispersion of spherical particles, depending on their hard-core diameter and strength of the flow field. As already alluded to, the effect of flow on the percolation threshold is modest with deviations of at most 15\%. See also the insets of Fig. \ref{fig:percflow}, that show relative changes compared to the no-flow case.

For all P\'{e}clet numbers studied, we find that there is a value of $D/\lambda_c$, below which the percolation threshold  increases and above which it decreases if compared with the $\mathrm{Pe}=0$ case. Both the decrease and increase are explained by the theory. Since the orientationally averaged correction of the pair correlation function is positive for short distances, see Fig.~\ref{fig:fullplotsimlines}(f), the flow field induces an increase in the average number of neighbours for small $\lambda$. This naturally translates to a lower percolation threshold for large $D/\lambda$. Conversely, for large separation distances $r$, the orientationally averaged correction becomes negative, meaning that the flow field decreases the number of neighbours for large $\lambda$, thereby increasing the percolation threshold for small $D/\lambda$.

We believe that the shear-induced decrease of the percolation threshold for large $D/\lambda_c$ is closely related to the emergence of so-called shear-induced contact clusters \cite{de1979conjectures}, which have also been found in earlier simulations \cite{gallier2015percolation,thogersen2016transient}. Even at volume fractions where such clusters are finite, they might play a significant role in aiding long-range connectivity and therefore in decreasing the percolation threshold with respect to the equilibrium situation. In fact, from our simulations we find that the average cluster becomes progressively elongated as the P\'eclet number increases, in line with what has been found in simulations of such contact clusters \cite{gallier2015percolation}.

The results of Fig. \ref{fig:percflow} confirm that there is good qualitative agreement between our theoretical model and simulation results of the percolation threshold. Our model correctly predicts the shear flow to induce an increase and subsequent decrease of the percolation threshold with decreasing connectivity range, and gives an accurate estimate of the location of the crossovers between these two regimes. It predicts the shift in the percolation threshold with almost quantitative accuracy as long as the material is sufficiently dilute, that is, as long as, say, $D/\lambda_c<0.8$, which roughly corresponds to hard-core volume fractions $\varphi<0.2$. 
However, as seen more clearly in the insets of Fig. \ref{fig:percflow}, our theory seems does not quite capture the impact of shear flow on the percolation threshold for large values of $D/\lambda_c$. In that case, the hard-core volume fraction is high and the shear flow significantly affects many-body contributions to the pair structure, as we also show in Fig. \ref{fig:fullplotsim}.  
The reason why a flow field seems to only have a modest effect on the percolation threshold for all values of $D/\lambda_c$, even though it strongly impacts on the pair structure, is probably due to the circumstance that the orientational average of the many-body corrections remains small. This can in fact be deduced from Figs. \ref{fig:fullplotsim}(c, f), which show that corrections in the compression quadrants in part compensate for those those in the extensional quadrants, especially in the low P\'{e}clet number regime. 

We expect that by taking many-body correlations in the distortion of the pair correlation function by the flow field explicitly into account, the accuracy of our predictions would improve slightly for values of $D/\lambda_c$ approaching unity. Fortunately, for the largest part of our work, the hard-core volume fractions at the percolation threshold remain rather low, that is, lower than 0.2 as long as the hard-core diameter remains smaller than 80\% of the connectivity range. At least for those cases, many-body correlations play only a subdominant role \cite{yurkovetsky2006triplet}, and should not be expected to influence the results much, as is evidenced by the agreement between the percolation thresholds obtained from our theory and simulations.  

Even if many-body correlations had been included, they would not have remedied the slight disagreement between theory and simulation of the location of the minimum of the percolation threshold as a function of $D/\lambda_c$ that the theory overestimates. As this is already the case for $\mathrm{Pe}=0$, this must in part be caused by the heuristic percolation theory we use. It would be interesting to see whether a novel continuum percolation theory that was recently proposed, nearest-neighbour connectedness theory \cite{coupette2020continuum,coupette2021nearest}, improves the agreement. Although it is also based on geometric considerations, it is not obvious how to apply it to out-of-equilibrium percolation as it does not rely solely on the concept of a pair correlation function.

For reasons of conciseness, neither our theory nor our simulations include hydrodynamic interactions between the colloids, e.g., in the form of short-ranged lubrication forces or long-ranged hydrodynamic many-body interactions. For low volume fractions and small P\'{e}clet numbers, we expect that changes in the perturbation of the structure due to hydrodynamic interactions are of quantitative nature only, and consequently this must also hold for their impact on the percolation threshold. In the limiting case of vanishing flow fields, indeed, the material is in equilibrium and hydrodynamic interactions cannot change the structure of the dispersion \cite{dhont1996introduction}. Consequently, they can neither have an effect on the percolation threshold. For strong shear cases, their exclusion might not be justified.

\section*{Conclusions and Outlook}\label{sec5}

We have presented a theoretical framework that describes the geometric percolation of hard, spherical particles under shear flow within the free-draining approximation. The predictions of the theory compare favourably with results of our Langevin dynamics computer simulations. We find that the percolation threshold is determined, on the one hand, by the ratio of the connectivity range and the hard-core diameter of the particles and, on the other, by the P\'{e}clet number that measures how much diffusion is affected by the flow field. The P\'{e}clet number is proportional to the flow rate. 

In the absence of a flow field, so at zero P\'{e}clet number, and at fixed connectivity range, the percolation threshold initially  decreases with increasing hard-core diameter, to subsequently increase again when the hard-core diameter approaches the connectivity range. If we ramp up the strength of the flow field, the percolation threshold \textit{increases} for small hard-core diameters smaller than some critical value, yet \textit{decreases} if the diameter is larger than that. This critical value we find to depend on the P\'{e}clet number. Our calculations show that this is caused by the balance between compressional and extensional effects that the shear flow exerts on the local particle density field near a reference particle.

According to our simulations and theoretical predictions, the influence of simple shear flow on the percolation threshold of hard colloidal spheres remains modest for P\'{e}clet numbers below 10. This suggests that for many practical applications, where a high-precision prediction is not required, the subtle influence of shear flow on the percolation threshold may well be disregarded. In that case, calculations based on equilibrium connectedness percolation theory would suffice. Whether this conclusion extends to sticky colloids or to particles that are not isometric, remains to be seen. 

The disagreement of our theoretical predictions with our computer simulations of the percolation threshold is at most 7\% for P\'{e}clet numbers up to ten. This error depends only very weakly on the shear rate, and because of that must originate mainly from either the heuristic percolation criterion or the equilibrium pair correlation function. It becomes shear rate dependent only if the hard-core diameter approaches the connectivity range. In that case, there are obvious routes to improvement for the theory. A clear path forward involves the inclusion of higher-than-two-body correlations in the bare correlation that serves as input for the calculations. See, for example, Refs \cite{lionberger1997smoluchowski, dhont1996introduction, hansen, szamel1992long, wagner2001smoluchowski, nazockdast2012microstructural} for more detailed discussions on how to deal with the inclusion of such higher order correlation functions.

Throughout this work, we have disregarded hydrodynamic interactions because they would render the molecular dynamics simulations very computationally expensive, and complicate the analytic treatment of the theory. As we argue in the main text, we expect that this does not introduce serious errors to the percolation threshold for low P\'{e}clet numbers. For stronger flow fields, however, hydrodynamic interactions should probably be taken into account to retain qualitative accuracy of the theory \cite{nazockdast2012microstructural, morris2009review}. We note that this is highly non-trivial, and requires the inclusion of additional terms in equation \eqref{eq:smol} \cite{lionberger1997smoluchowski,brady2001computer, vermant2005flow, dhont1989distortion, morris2009review, morris2002microstructure, tomilov2013aggregation, ball1995lubrication}.

Because of its simplicity and generality, the framework outlined here may prove a useful tool for modelling percolation in more realistic and perhaps interesting systems or setups. 
Our theory can straightforwardly be extended to more complicated flow fields and interaction potentials by solving the two-body Smoluchowski equation \eqref{eq:smol} either analytically or numerically, and inserting the result in the heuristic percolation criterion.

To apply the theory to the case of non-spherical particles, the pair correlation function depends also on the orientations of both particles and additional terms should be added in equation \eqref{eq:smol} to account for rotational diffusion \cite{tao2006isotropic, dhont1996introduction}. Computer simulations and experiments indicate that such orientational degrees of freedom give rise to fundamentally different clustering behaviour than is the case for dispersions of spherical particles \cite{kwon2012electrical, eken2011simulation, xu2006shear, alig2007electrical, bauhofer2009review}. It appears this is due, in large part, to the aligning effect a flow field exerts on highly non-isometric particles \cite{dhont2003viscoelasticity, ripoll2008attractive}.

To study the percolation of colloidal particles in polymeric hosts, which is the original motivation of this work, the effects of the colloidal attractions should be incorporated into the theory. We intend to pursue this in further research. Experiments suggest that these effects have a major influence on the percolation threshold of the nanocomposite \cite{schueler1997agglomeration}.

\section*{Methods}
{\small
\subsubsection*{Analytical solution of the Smoluchowski Equation}
We follow B\l{}awzdziewicz and Szamel to solve the two-particle Smoluchowski boundary value problem given by Eqs. (\ref{pde}) and (\ref{bcs}), using the method of induced multipole sources \cite{blawzdziewicz1993structure, duffy2015green}. First, 
we note that the fundamental solutions of the time-dependent version of the Smoluchowski equation \eqref{pde} 
\begin{equation}\label{eq:smolt}
    \frac{\partial \delta g}{\partial t} + \dot{\gamma}y\frac{\partial \delta g}{\partial x}-2D_0\nabla^2 \delta g=0
\end{equation}
are given by \cite{elrick1962source}
\begin{equation}
    F(x,y,z,t) = \frac{t^{-3/2}}{\sqrt{\frac{1}{3}t^2+1}}\exp\left[-\frac{\mathrm{Pe}}{4tD^2}\left(\frac{(x-yt)^2}{\frac{1}{3}t^2+1} + y^2+z^2\right)\right].
\end{equation}
To be able to satisfy the boundary conditions \eqref{bcs}, we place a linear combination of multipole sources at $\textbf{r}=0$. The multipole source functions $T_{\alpha \beta}(x,y,z, t)$ can be found by taking repeated spatial derivatives of the fundamental solutions
\begin{equation}
 T_{\alpha \beta}(x,y,z,t) = \left(D\frac{\partial}{\partial x}\right)^\alpha\left(D\frac{\partial}{\partial y}+2tD\frac{\partial }{\partial x}\right)^\beta F(x,y,z,t).
\end{equation}
Since we are focused on finding the steady-state solution of equation \eqref{eq:smolt} rather than its time-dependent behaviour, we use the limit method \cite{cole2010heat} to find steady state multipole source functions $T_{\alpha\beta}(x,y,z)$
\begin{equation}
    T_{\alpha\beta}(x,y,z) = \int_0^\infty \mathrm{d}t\, T_{\alpha\beta}(x,y,z, t).
\end{equation}

The full solution of the steady state boundary value problem is now
\begin{equation}\label{gTab}
\delta g(\textbf{r}, \mathrm{Pe}) = \sum_{\alpha, \beta=0}^\infty C^{\alpha\beta}  T_{\alpha \beta}(x,y,z), 
\end{equation}
in which $C^{\alpha\beta}$ are coefficients that are used to satisfy the boundary condition \eqref{bcs}. 

To fix the coefficients $C^{\alpha\beta}$, we insert \eqref{gTab} into \eqref{bcs} and expand both the left- and the right-hand side in terms of real-valued spherical harmonics $Y_{lm}(x,y,z)$. Using the fact that $xy=D^2\sqrt{4\pi/15}Y_{2,-2}(x,y,z)$, the boundary condition can now be formulated as
\begin{equation}
    \sum_{\alpha,\beta=0}^\infty C^{\alpha\beta}\sum_{l=0}^\infty\sum_{m=-l}^l j_{\alpha\beta}^{lm} Y_{lm}(x,y,z) = \sqrt{\frac{4\pi D^2}{15}}Y_{2, -2}, \qquad r=D,
\end{equation}
or, more compactly,
\begin{equation}\label{linsys}
    \sum_{\alpha,\beta=0}^\infty C^{\alpha\beta}j_{\alpha\beta}^{lm} = \sqrt{\frac{4\pi D^2}{15}} \delta_{l,2}\delta_{m,-2},
\end{equation}
from which the expansion coefficients $j_{\alpha\beta}^{lm}$ follow by invoking the orthogonality property
\begin{equation}
    j_{\alpha\beta}^{lm} = \int_0^{2\pi}\mathrm{d}\phi\int_0^\pi\mathrm{d}\theta\sin\theta Y_{lm}(\theta,\phi)\left(\left.\frac{\partial T_{\alpha\beta}}{\partial r}\right.-\frac{xy}{D}T_{\alpha\beta}(x,y,z)\right),
\end{equation}
evaluated at $r=D$. Having found the expansion coefficients $j_{\alpha\beta}^{lm}$, we evaluate the coefficients $C_{\alpha\beta}$ by inverting the linear system of equations given by \eqref{linsys}. 
In agreement with B\l{}awzdziewicz and Szamel, we find that the procedure converges if we take into account all coefficients such that $\alpha+\beta\leq10$, at least for $\mathrm{Pe}<10$. 

The spherical integrals we performed numerically using the Lebedev quadrature of order 47 \cite{lebedev1976quadratures}. The $t$-integrals we performed using Simpson's rule on a logarithmic grid of 100 points spanning from $t=10^{-5}$ to $t=10^5$ \cite{pang1999introduction}. In order to calculate the radial integrals occurring in the expressions of the two length scales $l$ and $L$, we use a trapezoidal rule, and subsequently solve the percolation criterion $2l=L$ using Broyden's rule \cite{broyden1965class}, in order to find the percolation threshold.

\subsubsection*{Molecular dynamics simulations}
In order to verify our theory, we perform particle-resolved simulations using the LAMMPS software package \cite{thompson2021lammps}, explicitly integrating the Langevin equation 
\begin{equation}
    m\ddot{\textbf{r}} = -\nabla U - \gamma \dot{\textbf{r}} + \textbf{R},
\end{equation}
for every particle \cite{doi2013soft}. In the Langevin equation, we have introduced the mass $m=1$, friction coefficient $\gamma=10$, and the fluctuating force $\textbf{R}(t)$ that has zero mean and a variance dictated by the fluctuation dissipation theorem
\begin{equation}
    \left<R_i(t)R_j(t')\right> = 2\gamma k_BT \delta_{i,j}\delta(t-t'),
\end{equation}
in which $R_i$ and $R_j$ are components of the vector $\textbf{R}$. The potential energy is given by the sum over all pair potentials, for which we choose the Weeks-Chandler-Anderson form \cite{weeks1971role},
\begin{equation}
    U_{ij}(r)/k_BT = 4\epsilon\left(\left(\frac{\sigma}{r_{ij}}\right)^{12}-\left(\frac{\sigma}{r_{ij}}\right)^6\right)+\epsilon,
\end{equation}
if $r_{ij}<2^{1/6}\sigma$ and $U_{ij}/k_BT=0$ otherwise. We choose to set $\sigma=2^{-1/6}D$ and $\epsilon=100$. This choice ensures that we accurately model hard-sphere behaviour. 
We introduce shear flow in our simulations by deforming our simulation box every time step such that the shear strain remains equal to $\dot\gamma=1$ in dimensionless units. The P\'{e}clet number $\mathrm{Pe} = \frac{\dot\gamma D^2}{4D_0}$ we then vary through the diffusion constant $D_0=k_BT/\gamma$ by adjusting the temperature. 
By changing the density, we effectively change $D/\lambda_c$, allowing us to probe the influence of hard-core interactions on the percolation threshold. 

We set our time step equal to $\Delta t=10^{-3}$ and perform production runs of $10^8$ time steps, saving the particle positions every $10^4$ time steps. By performing a full hierarchical clustering procedure for each saved simulation snapshot \cite{johnson1967hierarchical, mullner2013fastcluster}, we determine the smallest connectivity length $\lambda$ for which a cluster exists that connects to any image of itself through the periodic boundaries \cite{rapaport2004art}. Using the set of all such connectivity lengths, we construct a percolation probability $P(\lambda)$ for a given simulation. An example of such curves is given for four different system sizes in Fig.~\ref{fig:fullplotsimlines}(h).

To obtain the percolation threshold $\lambda_c$, we locate of the intersection of the percolation probabilities of simulations of two different system sizes at the same particle density \cite{vskvor2007percolation}. 
We find that finite-size effects are negligible if the sizes of the two systems are chosen such that $N=500$ and $N=4000$. In the absence of a shear flow $\dot\gamma = 0$, our obtained percolation thresholds agree to within a few percent with those found in the literature \cite{miller2009structural}. We probe the effect of the flow field on cluster shape by evaluating the eigenvalues of the gyration tensors of clusters of connected particles \cite{sanchez2005equilibrium}. 

}

\section*{Acknowledgments}

Ilian Pihlajamaa has been financially supported by the Dutch Research Council (NWO) through a Vidi grant, and Ren\'e de Bruijn and Paul van der Schoot acknowledge funding by the Institute for Complex Molecular Systems at Eindhoven University of Technology. 

\section*{Code availability} The code that is used to produce all data presented in this work is available in a permanent repository with DOI: \url{https://doi.org/10.5281/zenodo.5786307}.
\end{multicols}

\bibliography{bibliography.bib}

\end{document}